\documentstyle[aps,epsf,preprint]{revtex}
\newcommand{\Ta}{H_T}
\newcommand{\Tb}{\tilde{H}_T}
\newcommand{\Tc}{E_T}
\newcommand{\Td}{\tilde{E}_T}

\def\la{\langle}\def\ra{\rangle}

\def\roughly#1{\mathrel{\raise.3ex\hbox{$#1$\kern-.75em
\lower1ex\hbox{$\sim$}}}}

\begin{document}

\preprint{\hfill\parbox[b]{0.3\hsize}
{ }}

\def\bra{\langle }
\def\ket{\rangle }

\title{Model study of 
generalized parton distributions with helicity flip}

\author
{
Sergio Scopetta}

\address
{\it 
%(a)
Dipartimento di Fisica, Universit\`a degli Studi
di Perugia, 
and INFN, sezione di Perugia
\\
via A. Pascoli
06100 Perugia, Italy, 
}
%\\
%(b)
%Departament de Fisica Te\`orica,
%Universitat de Val\`encia, 46100 Burjassot (Val\`encia), Spain
%\\
%and Institut de F\'{\i}sica Corpuscular,
%Consejo Superior de Investigaciones Cient\'{\i}ficas }

\maketitle

\begin{abstract}
Generalized parton distributions with helicity flip
are studied in the quark sector, within a simple version 
of the MIT bag model, assuming an SU(6) wave function 
for the proton target. In the framework under scrutiny it turns
out that only the generalized transversity distribution,
$H_T^q$, is non vanishing. For this quantity, the forward limit 
is properly recovered and numerical results are found to
underestimate recent lattice data for its first moment. 
Positivity bounds
recently proposed are fulfilled by the obtained distribution.
The relevance of the analysis for the planning of 
measurements of the quark generalized transversity
is addressed.
\end{abstract}
%\noindent{\small$\dagger$

%\section{Introduction}

The distribution of transverse quark spin  
is one of the least known features of nucleon structure. In a few years,
some light should be shed on it by experiments
\cite{trans}. The possibility of measuring the generalized
transverse spin distribution is also under scrutiny,
establishing a link between transversity and 
Generalized Parton Distributions (GPDs) \cite{first}.
GPDs represent one of the main topics of interest in nowadays 
hadronic physics \cite{md}.
At twist-two, eight GPDs occur.
Four of them, helicity conserving ones,
enter processes where
{\sl the helicity of the parton is conserved}.
They are labelled $H,E,\tilde H, \tilde E$
and have been extensively studied and modelled.
The other four twist-two GPDs,
$H_T,E_T,\tilde H_T, \tilde E_T$,
the subject of this study,
are {\sl parton helicity flip} ones
and have been introduced in Ref. \cite{hod},
although their correct classification and counting
have been established later, in Ref. \cite{die1}.
Being diagonal in a transversity basis, they
are also called ``transversity GPDs''.
I prefer to call them ``GPDs with helicity flip'',
calling generalized transversity distribution the 
quantity $H_T$, the only one
which survives in the forward limit, yielding the 
transversity density, $h_1$.
While gluon helicity flip GPDs  
appear at leading twist in Deeply Virtual Compton Scattering
\cite{bmk}, the same does not occur in the quark sector,
not even in hard exclusive electroproduction
of mesons \cite{pi1}.
Diffractive double
meson production is
the only process which is known to give access
to the quark generalized transversity, in the 
Efremov-Radyushkin-Brodsky-Lepage (ERBL) region
\cite{pire}.
An estimate of helicity flip GPDs would help
to study the feasibility of such an experiment.
Recently, lattice data 
for their lowest moments 
\cite{lat} and a study of positivity bounds
on them have appeared \cite{goe}.
Studies of helicity flip GPDs in the transverse plane
have been completed \cite{die2}.
Anyway, to my knowledge, a model calculation of these quantities 
has not been performed yet.  
In here, a model estimate of 
quark GPDs with helicity flip 
is presented.
The analysis is performed within the MIT bag model
\cite{MIT},
assuming SU(6) symmetry, 
following the lines of Ref. \cite{ji1},
where helicity conserving GPDs have been calculated.
Despite of well known drawbacks,  
such as the breaking of translational invariance, 
%rather thorny for evaluating form factors and GPDs,
the MIT bag model has proven to be able
to provide reasonable initial inputs, 
at a low factorization scale, for the unpolarized \cite{jaf}, 
polarized \cite{jija1}, transversity \cite{jija2,strat} and orbital
angular momentum \cite{sv2} distributions.
The MIT bag model has been also the framework
for the first estimate of helicity conserving
GPDs \cite{ji1}, up to twist three
\cite{dan},
and it represents therefore the natural playground
for the first analysis of the helicity flip ones.
\vskip 2mm

%\section{GPDs with helicity flip}
The main quantities of interest are now defined.
Quark helicity flip GPDs, 
%$H_T^q,E_T^q,\tilde H_T^q, \tilde E_T^q$,
are introduced through the relation 
\cite{die1} 
\begin{eqnarray}
  \label{flip-quark}
\lefteqn{ \frac{1}{2} \int \frac{d z^-}{2\pi}\, e^{ix P^+ z^-}
  \langle p',\lambda'|\, 
     \bar{\psi}_q(-{\textstyle\frac{1}{2}}z)\, i \sigma^{+i}\, 
     \psi_q({\textstyle\frac{1}{2}}z)\, 
  \,|p,\lambda \rangle \Big|_{z^+=0,\, \mathbf{z}_T=0} } 
\nonumber \\
&=& \frac{1}{2P^+} \bar{u}(p',\lambda') \left[
  \Ta^q\, i \sigma^{+i} +
  \Tb^q\, \frac{P^+ \Delta^i - \Delta^+ P^i}{m^2} \right.
\nonumber \\
&& \left. \hspace{5em} {}+
  \Tc^q\, \frac{\gamma^+ \Delta^i - \Delta^+ \gamma^i}{2m} +
  \Td^q\, \frac{\gamma^+ P^i - P^+ \gamma^i}{m}
  \right] u(p,\lambda)~,
\hspace{2em}
\end{eqnarray}
where 
$p, p'$ and $\lambda, \lambda'$ respectively denote 
momenta and helicities
of the nucleon and $i=1,2$ is a transverse index.
Use is made of light-cone coordinates ($v^\pm = (v^0
\pm v^3)/\sqrt{2}$ and ${\mathbf{v}_T} = (v^1, v^2)$ for any four-vector
$v$) and of Ji's kinematical variables,
%\begin{equation}
%  \label{kin-def}
$
P = (p+p')/2 ,
\Delta = p'-p , 
\xi= ({p^+-p'^+})/({p^+ +p'^+})=- {\Delta^+}/{2P^+} ,
%\end{equation}
$
and $t = \Delta^2$. 
It is convenient to use light-cone
helicity states and to have the quarks on shell, so that
the operators occurring in the definitions of the quark
distributions have the simplest structure. With this choice
the helicity flip GPDs turn out to be related to the matrix elements
$A_{\lambda'\mu',\lambda \mu}$, for definite parton
helicities $\mu'=+$ and $\mu=-$ 
\begin{eqnarray}
  \label{on-shell}
A^q_{\lambda'+, \lambda -} &=&
\int \frac{d z^-}{2\pi}\, e^{ix P^+ z^-}
  \langle p',\lambda'|\, {\cal O}^q_{+,-}(z)
  \,|p,\lambda \rangle \Big|_{z^+=0,\, \mathbf{z}_T=0} 
\nonumber \\
&=& \int \frac{d^2 k_T}{(2\pi)^3}
\left[\, \int dz^-\,d^2 z_T\, e^{i k\cdot z}\,
       \langle p',\lambda'|\, {\cal O}^q_{+,-}(z) \,
       |p,\lambda \rangle\,
     \right]_{z^+=0,\, k^+ = x P^+}~~,
\end{eqnarray}
with the operator $ {\cal O}_{+,-}^q(z) $, given by
\begin{eqnarray}
{\cal O}^q_{+,-}(z) = \frac{i}{4}\, 
\bar{\psi}_q\left (-{ z \over 2} \right ) \, (\sigma^{+1} + i \sigma^{+2})\, 
\psi_q \left ({z \over 2} \right)~,
\hspace{2em}
\end{eqnarray}
flipping the parton helicity from $\mu=-{1 \over 2}$ to $\mu'= {1 \over 2}$
\cite{die1}.
In a model study of GPDs with helicity flip, the crucial
calculation is therefore the evaluation
of the matrix elements Eq. (\ref{on-shell}).
Once these results are available, the GPDs are obtained from them
and their explicit form can be found, i.e., in Ref. \cite{goe}.
%
%\begin{eqnarray}
%  \label{flip-inv}
%\Ta^q &=& { 1 \over \sqrt{1 - \xi^2} }
%\left \{ A_{++,--}^q + A_{-+,+-}^q - { \xi \over \sqrt{1 - \xi^2} }
%{a \over \epsilon} ( A_{++,+-}^q - A_{-+,--}^q ) \right \}~,
%\nonumber \\
%\Tb^q &=& - { a^2 \over \sqrt{1 - \xi^2} } A_{-+,+-}^q~,
%\nonumber \\
%\Tc^q &=& { a \over {1 - \xi}}
%\left \{ { 1 \over \epsilon}
%\left ( { 3 + \xi \over 1 + \xi }  A_{-+,--}^q - A_{++,+-}^q \right )
%+ {2a \over \sqrt{1 - \xi^2}} 
%{A_{-+,+-}^q  \over 1 + \xi}
%\right \}~,
%\nonumber \\
%\Td^q &=& { a \over \epsilon } ( A_{++,+-}^q - A_{-+,--}^q ) +
%{ \xi a \over {1 - \xi}}
%\left \{ { 1 \over \epsilon}
%\left ( { 3 + \xi \over 1 + \xi }  A_{-+,--}^q - A_{++,+-}^q \right )
%+ {2a \over \sqrt{1 - \xi^2}} 
%A_{-+,+-}^q
%\right \}~,
%\end{eqnarray}
%where $ a = 2m / \sqrt{t_0 - t}$, 
%$t_0 = - {4 m^2 \xi^2}/({1-\xi^2})$
%is the maximum value of $t$ for given $\xi$, and $\epsilon =
%\mathrm{sgn}(D^1)$, where $D^1$ is the $x$-component of $D^\alpha =
%P^+ \Delta^\alpha - \Delta^+ P^\alpha$. The case $D^1 = 0$ corresponds
%to $t=t_0$.
%\section{MIT bag model estimates of GPDs}
\vskip 2mm
The procedure of Ref. \cite{ji1}
for estimating GPDs is adopted here,
using a simple version
of the MIT bag model, able to
reproduce the gross features of parton distributions
\cite{jaf,jija1,jija2,strat,sv2} 
and form factors (ff)
\cite{BETZ}, despite of its drawbacks later discussed.
When evaluating GPDs it is convenient to work in the
Breit frame, where 
$p_{\mu} = ( \overline m; -\overrightarrow \Delta/2 )$,
$p'_{\mu}\ =\ ( \overline m; \overrightarrow \Delta/2 )$,
$
t\ = \Delta^2 = \ - \overrightarrow\Delta^2
= 4 \left( m^2 - \overline m^2 \right)
$
%%
%Using Eqs.(2), we then have:
%%
%\begin{eqnarray}
%p^{\mu} &=& (1; 0,0,1)/(2\overline m),\ \ \ \ \ \ \
%n^\mu = (1; 0, 0, -1)/\overline{m}.
%\end{eqnarray}
%
and 
$
\xi\ =\ - \Delta_z / ( 2 \overline m ).
$
In principle,
since translational invariance is violated,
different results will be obtained in
different frames.
As in Ref \cite{ji1},
it is assumed here that the results are
weakly frame dependent.
The calculation, performed for
quarks of minimum energy
in the bag, requires wave functions of moving nucleons
and one has to boost the rest frame wave function
%$\psi(t, \vec r)$,
to a moving frame
\cite{BETZ}.
Here, following Ref. \cite{ji1},
a simple prescription is used to partially restore momentum conservation,
taking the momentum transfer through the active quark
to be $\eta \vec \Delta$,
where $\eta$ is a parameter to be fixed by fitting the 
nucleon electromagnetic ff
\cite{BETZ}. 
In Ref. \cite{ji1},
it was found that small $|t|$ data
favor a value of $\eta=0.55$, while a better fit is achieved at larger
$|t|$ with $\eta=0.35$.
\vskip 2mm
Following this approach, I obtained
the following expression for the matrix elements
Eq. (\ref{on-shell}):
\begin{eqnarray}
\label{H}
A^q_{\lambda'+, \lambda -}(x,\xi,t) 		
& = & 
\sqrt{1 -\xi^2}
Z^2(t) \left(4 \pi N^2 R^6\right)
{ \overline{m} \over 1 - C_1 \Delta_z^2/t }
\int { dk_\perp\ d\varphi \over (2\pi)^3 }\ k_\perp		\nonumber\\
& \times &
\left\{ 
\ C \ t_0(k)\ t_0(k')
+\
\ C \ 
{k_z \over k }
t_1(k)\ t_0(k') 		
\right.					\nonumber\\
& + &
\left. 
\left[
\ C \ + 2 \ { \Delta_x \widetilde \Delta_x \over |\vec \Delta| k_z'}  
\left ( \cosh {\omega\over 2} \sinh {\omega\over 2}
- {\Delta_z \over  |\vec \Delta | }  
  \sinh^2{\omega\over 2} \right )
\right ]
{ k_z' \over k' }
t_0(k)\ t_1(k') 
\right.					\nonumber\\
& + & 
\left.
\left[ \ C \ + \cosh {\omega\over 2} \sinh {\omega\over 2}
{ \Delta_x \widetilde \Delta_x \over |\vec \Delta| k_z'}
+
{ \Delta_x \widetilde \Delta_x \Delta_z \over t k_z'}
\right]
{k_z k_z' \over k k' }
t_1(k)\ t_1(k') \right\}
\nonumber \\
& \times &
\la \ p'\ \lambda'\ | \ b^{\, q \ \dagger}_+ \ b^{\ q}_- 
\ | \ p \ \lambda \ \ra~,
\end{eqnarray}
where 
$
%\begin{equation}
%\label{C}
C = \cosh ^2 ({\omega / 2})
+  (\Delta_z^2 /t) \sinh ^2 (\omega/2)~,
%\end{equation}
k' \equiv |\vec k'|$,
$\vec k' = \vec k + \overrightarrow{\widetilde{\Delta}}$ and
the effective momentum transfer is
%
% \begin{eqnarray}
$
\overrightarrow{\widetilde\Delta}
= \eta\ { \overrightarrow{\Delta} / \cosh\omega }.
%\end{eqnarray}
$
For simplicity, 
it has been chosen 
$\vec \Delta = (\Delta_x,0,\Delta_z)$.
The explicit forms of the
spectator term $Z(t)$, of the functions $t_0(k)$ and $t_1(k)$,
of the normalization $N$ and of $k_z$
are given in \cite{ji1}.
%so that $k_z$, giving the $x-$dependence in Eq. (\ref{H}), reads
%%
%\begin{eqnarray}
%k_z =
%{ \overline m \over 1 - 
%%(\cosh\omega-1)
%C_1 
%\Delta_z^2/t }		
%%\nonumber\\
%% &\times&
%\left[ x - {1 \over 2 \overline m}
%   \left( (2 \epsilon_0 + \widetilde\Delta_z) \cosh\omega
%	+ |\overrightarrow{\widetilde\Delta}| \sinh\omega
%	- C_1 {2 \Delta_x \Delta_z \over t } k_x 
%%(\cosh\omega-1)
%   \right)
%\right]~.
%\end{eqnarray}
%%
Besides, in Eq (\ref{H}), 
$ \cosh\omega = { \overline{m} / m }$,  
%~~~~~~
$\sinh\omega = { |\overrightarrow\Delta| / (2 m) }$,
$R$ is the bag radius,
related to the quark energy
$\epsilon_0 = \omega_0/R$, being
$\omega_0 = 2.04$ the lowest frequency solution of the bag eigenequation, 
given in turn by the relation:\ $R m = 4 \omega_0$ \cite{MIT,jija1}.
If $SU(6)$ symmetry is assumed,
as it has been done in Ref. \cite{jija1,jija2,strat}
for bag model calculations of parton
distributions, or in Ref. \cite{ji1} for 
bag model calculations of helicity conserving GPDs,
the matrix elements 
$ 
\la \ p'\ \lambda'\ | \ b^{\, q \ \dagger}_+ \ b^{\ q}_- 
\ | \ p \ \lambda \ \ra
$, appearing in Eq. (\ref{H}), reduce to:
\begin{eqnarray}
\label{su6}
\la \ p'\ \lambda'\ | \ b^{\, q \ \dagger}_+ \ b^{\ q}_- 
\ | \ p \ \lambda \ \ra = \delta_{\lambda' +, \lambda -}
\end{eqnarray}
It turns out therefore that within the
MIT bag model in the lowest energy state in an SU(6) spin-flavour
scenario, among the helicity flip amplitudes Eq. (\ref{on-shell}),
$A_{++,--}^q$ is the only non vanishing one.
This is understood in terms of angular momentum conservation:
in order to flip the helicity of the quark keeping fixed
the one of the target, one has to assume target
{\sl orbital} angular momentum excitation, 
impossible in a pure SU(6) scenario.
As a consequence, in the present scheme, 
from Eq. (\ref{on-shell}) one gets that
the generalized transversity distribution
\begin{equation}
\label{HSU6}
{\Ta^q}(x,\xi,\Delta^2)={ 1 \over \sqrt{1 - \xi^2} } 
\, A^q_{++,--}(x,\xi,\Delta^2)
\end{equation}
is the only non vanishing GPD with helicity flip.
\vskip 2mm
Numerical results for ${\Ta^q}(x,\xi,\Delta^2)$, Eq. (\ref{HSU6}),
evaluated using Eq. (\ref{H}) with the SU(6) condition
Eq. (\ref{su6}), are shown in Figs 1 to 3.
The results have to be ascribed to the low factorization scale,
$\mu_0$, corresponding to the model,
assumed to be $\mu_0 = 0.4$ GeV as in Ref. \cite{ji1},
although it is not possible to fix it from first principles.
In Fig. 1, the forward limit of Eq. (\ref{HSU6}),
${\Ta^q}(x,\xi=0,\Delta^2=0)$,
is shown.
As expected, it coincides with the result presented, for the 
transversity distribution
$h_1^q(x)$, in Ref. \cite{jija2}.
In Figs. 2 and 3, the full $x$ and $\xi$ dependences
predicted by Eq. (\ref{HSU6}) are shown for 
$\Delta^2 = -0.5$ GeV$^2$ and 
$\Delta^2 = -1.$ GeV$^2$, respectively.
The value of the parameter $\eta$, fixing the effective 
momentum transfer, has been taken to be 0.55, as done in
Ref. \cite{ji1} for presenting the results.
The SU(6) $u$ flavour distribution would be obtained
by multiplying these results by 4/3; 
the $d$ one by multiplying them by -1/3.
The main features of the results
are similar to those obtained for the helicity conserving
sector \cite{ji1}:
a) a strong $\Delta^2$ dependence mainly governed by the ff;
b) a weak $\xi$ dependence, although a mild shift of the peak
toward larger $x$ can be observed when $\xi$ increases;
c) a little contribution in the 
ERBL region ($-\xi \leq x \leq \xi$).
Some of these features 
%commenting on
%Figures 2 and 3, namely 
%the slow $\xi$ dependence of the present results,
%similar to those found in Ref. \cite{ji1},
%together with a $\Delta^2$ dependence mainly governed
%by the ff, 
may be artifacts of the model
under scrutiny. They could be due 
to the approximations used.
Indeed, in the parton helicity conserving sector,
some model studies 
%performed in the full
%DGLAP and ERBL regions 
brought to rather different conclusions,
in particular concerning
the slow $\xi$ dependence of the results together with a $\Delta^2$ 
dependence mainly governed
by the ff (see \cite{md} for a summary of results).
For an easy discussion I summarize
the approximations hidden in the present approach:
i) quarks are in the lowest energy state; 
ii) the role of antiquarks is disregarded;
iii) the dependence of the results on the
choice of the reference frame is supposed to be weak;
iv) a possible effect of the bag boundary has been
neglected; 
v) momentum conservation is only partially restored
by a prescription motivated in Ref. \cite{BETZ,ji1};
vi) the spin-flavor structure has been taken to be $SU(6)$.
The model can be enriched in different aspects removing part
of the approximations i)-vi), which could lead
to different $\xi$ and $\Delta^2$ behaviours.
Finite distributions ${\Tb^q},{\Tc^q},{\Td^q}$
can be obtained by relaxing the assumption i) and/or the assumption vi);
a stronger $\xi$ dependence
could be obtained by a more transparent prescription
for restoring momentum conservation (approximation v).
The present analysis has been motivated in part by the necessity
of calculating cross-sections for
the process studied in Ref. \cite{pire}, which
could give access to $H_T^q$.
In that case, the main 
contribution comes from the ERBL region.
It will be therefore very interesting to extend
the study to antiquark degrees of freedom,
relaxing the approximation ii).
Besides, to predict realistic cross sections,
one has to evolve these low-factorization scale results
to experimental scales, according to pQCD.
This procedure will 
produce an enhancement of the distribution in the ERBL
region. 
The outcome of this analysis is compared now
with a recent lattice calculation \cite{lat}.
The first moment of $H_T^q$
is the ``tensor form factor''
\begin{equation}
\label{tff}
A_{T10}^q(\Delta^2) = \int_{-1}^1 dx \ H^q_T(x,\xi,\Delta^2)~,
\end{equation}
yielding at $\Delta^2=0$ the quark tensor charge.
Lattice data for the tensor ff have been recently reported
in Ref. \cite{lat}, where a dipole fit to them has also been 
proposed.
Having no experimental data on this quantity at disposal,
in Fig. 4 I compare the isovector $u-d$ tensor ff
Eq. (\ref{tff}), obtained by integrating Eq. (\ref{HSU6}),
with the dipole fit of {\sl lattice} data provided in Ref. \cite{lat}.
It is seen that the results obtained with the choices 
$\eta = 0.35$ and $\eta = 0.55$ lie below the points 
corresponding to the fit.
In Ref. \cite{ji1}, the comparison of the MIT 
bag model calculation for the
electromagnetic ff with {\sl experimental} data
had given a different outcome.
In that case,
at large $- \Delta^2$, data
were underestimated
by the calculation with $\eta = 0.55$ and overestimated
by taking $\eta= 0.35$. 
In the figure, 
the calculated ff has been divided by 1.35, the
value of the
isovector tensor charge predicted by the MIT bag model
with SU(6) symmetry.
For this quantity, the same lattice calculation yields
the value 1.09.
Recently, positivity bounds on helicity flip GPDs
have been derived in Ref. \cite{goe}. The strongest
bound on $H_T^q$,
Eq. (5.1) of Ref. \cite{goe},
% reads:
%\begin{equation}
%\label{go}
%| H_T^q (x,\xi,\Delta^2)| \leq { 1 \over 4 \sqrt{1 - \xi^2}}
%\sum_{j=1,3} A_j~,
%\end{equation}
%where the RHS depends on the three 
%forward parton distributions
%of twist two, $f_1^q,g_1^q$ and $h_1^q$.
%The constraint Eq. (\ref{go}) 
%has been
%numerically checked,
%by using Eq. (\ref{HSU6}) in the L.H.S.
%and the known results \cite{jaf,jija1,jija2} for $f_1^q,g_1^q$
%and $h_1^q$ for evaluating the RHS.
%Eq. (\ref{go}) 
has been found
to be fulfilled by the MIT bag with SU(6) symmetry,
in any kinematical region.
\vskip 2mm
%\section{Conclusions}

In summary, a first calculation of quark helicity flip 
GPDs has been presented.
The analysis is motivated in part by the necessity of 
estimating cross sections for a
physical processes which has been proposed to access 
generalized transversity. 
As it has been done in the past to obtain first 
estimates of different parton distributions,
the MIT bag model has been chosen, adopting 
SU(6) symmetry.
As expected, in SU(6) only the generalized transversity distribution
is non vanishing; the forward limit is recovered and
the main features of the full result, at the 
low factorization scale of the model, are a weak $\xi$ dependence
and a little contribution in the ERBL region.
The output of the calculation 
underestimates lattice data
and fulfills
recently proposed positivity bounds.
This work represents a first step for
a full modelling of parton helicity flip GPDs in the quark
sector, relevant for phenomenological studies.
More realistic estimates will be obtained by relaxing
the SU(6) assumption, taking into account antiquark
degrees of freedom, 
implementing a better prescription for restoring
translational invariance, evolving the model results 
to experimental scales. 
This will permit to obtain more reliable results
in the ERBL region.

%\acknowledgements
\vskip 2mm
I am grateful to B. Pire, for suggesting me 
to start this investigation and for useful comments.
Enlightening discussions with V. Vento are gratefully acknowledged.
This work is supported in part by the Italian MIUR
through the PRIN ``Theoretical Studies of the Nucleus
and the Many Body Systems''.

\newpage
\appendixonfalse
\section*{Figure Captions}

\vspace{1em}\noindent
{\bf Fig. 1}:
The GPD ${\Ta^q}$, Eq. (\ref{HSU6}), in the forward limit $\xi=0,\Delta^2=0$,
giving the transversity distribution $h_1^q(x)$.

\vspace{1em}\noindent
{\bf Fig. 2}:
The $x$ and $\xi$ dependences of
the helicity flip GPD ${\Ta^q}$, Eq. (\ref{HSU6}),
for $\Delta^2=-0.5$ GeV$^2$.

\vspace{1em}\noindent
{\bf Fig. 3}:
As in Fig. 2, but for $\Delta^2=-1.$ GeV$^2$.

\vspace{1em}\noindent
{\bf Fig. 4}:
The isovector ($u-d$) tensor form factor, Eq. (\ref{tff}), evaluated
for $\eta=0.35$ (full) and $\eta=0.55$ (dashed), 
divided by the isovector tensor charge predicted by the
MIT bag model with SU(6) symmetry,
compared with the dipole fit to the lattice prediction given
in Ref \cite{lat} (dots).

\newpage

\begin{figure}[ht]
%\vspace{6.6cm}
\includegraphics{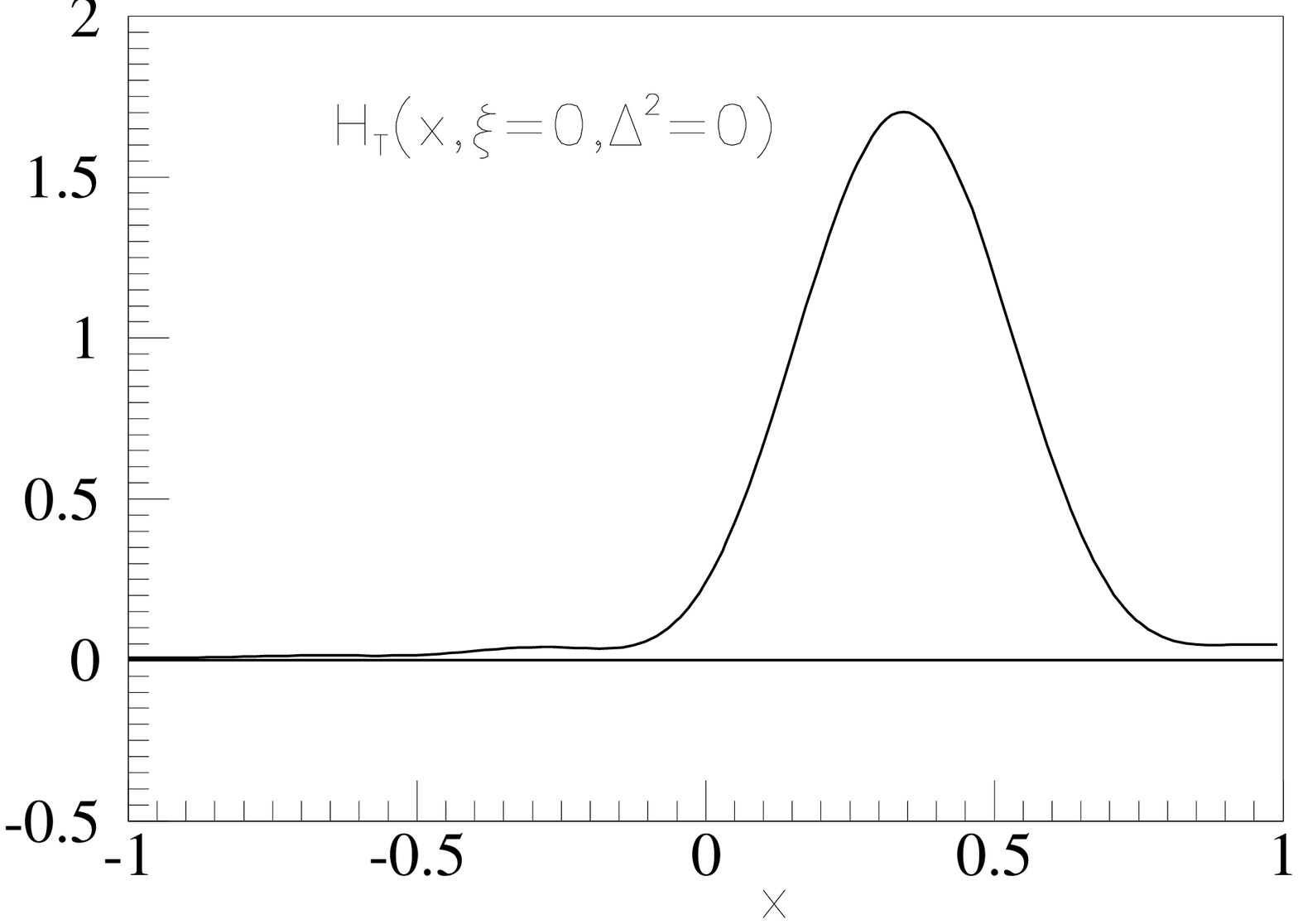}
\vspace{12.0cm}
\caption{}
\end{figure}

\newpage

\begin{figure}[h]
%\vspace{20.0cm}
\includegraphics{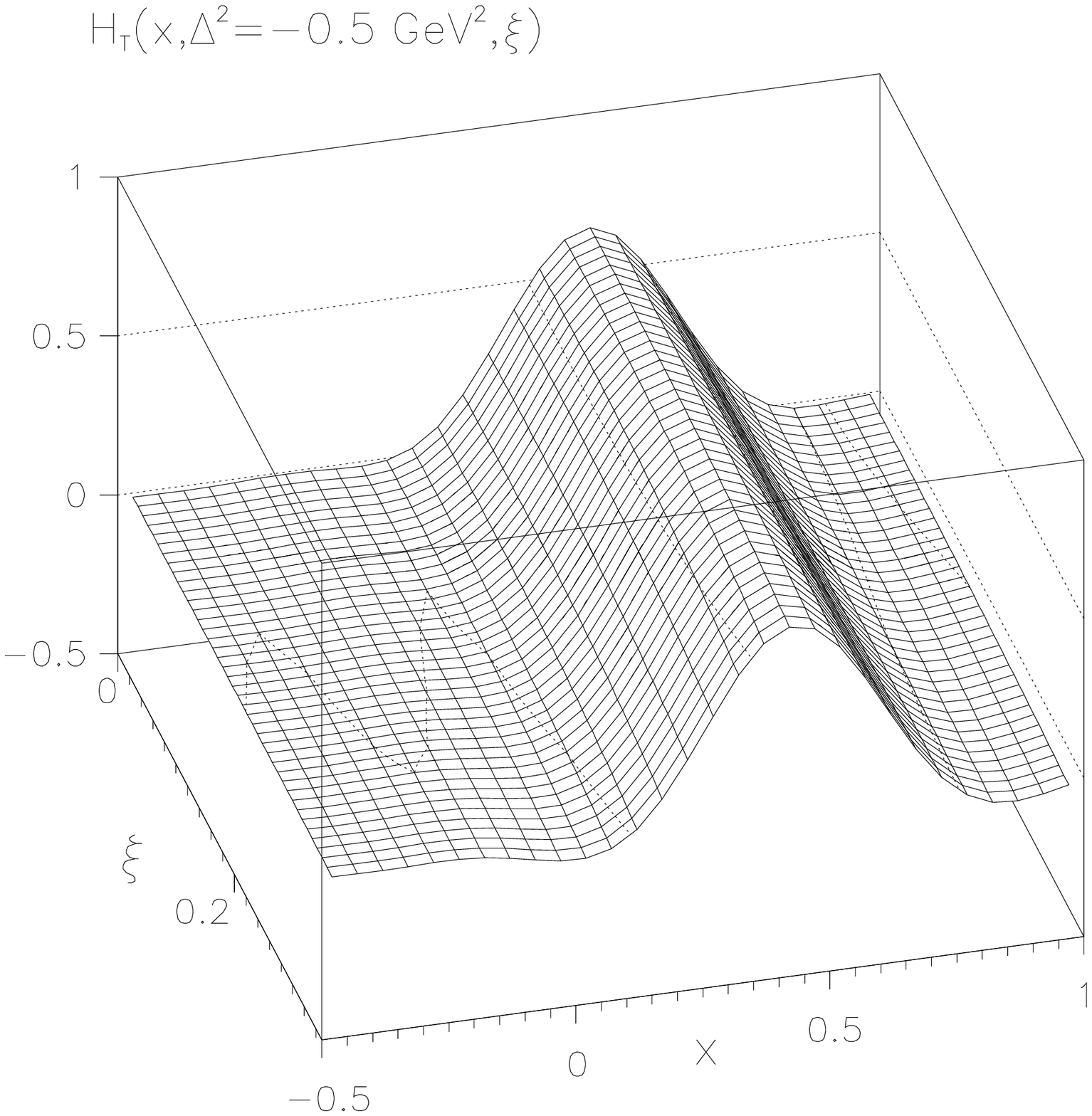}
\vspace{12.0cm}
\caption{}

\end{figure}

\newpage

\begin{figure}[h]
%\vspace{24.cm}
\includegraphics{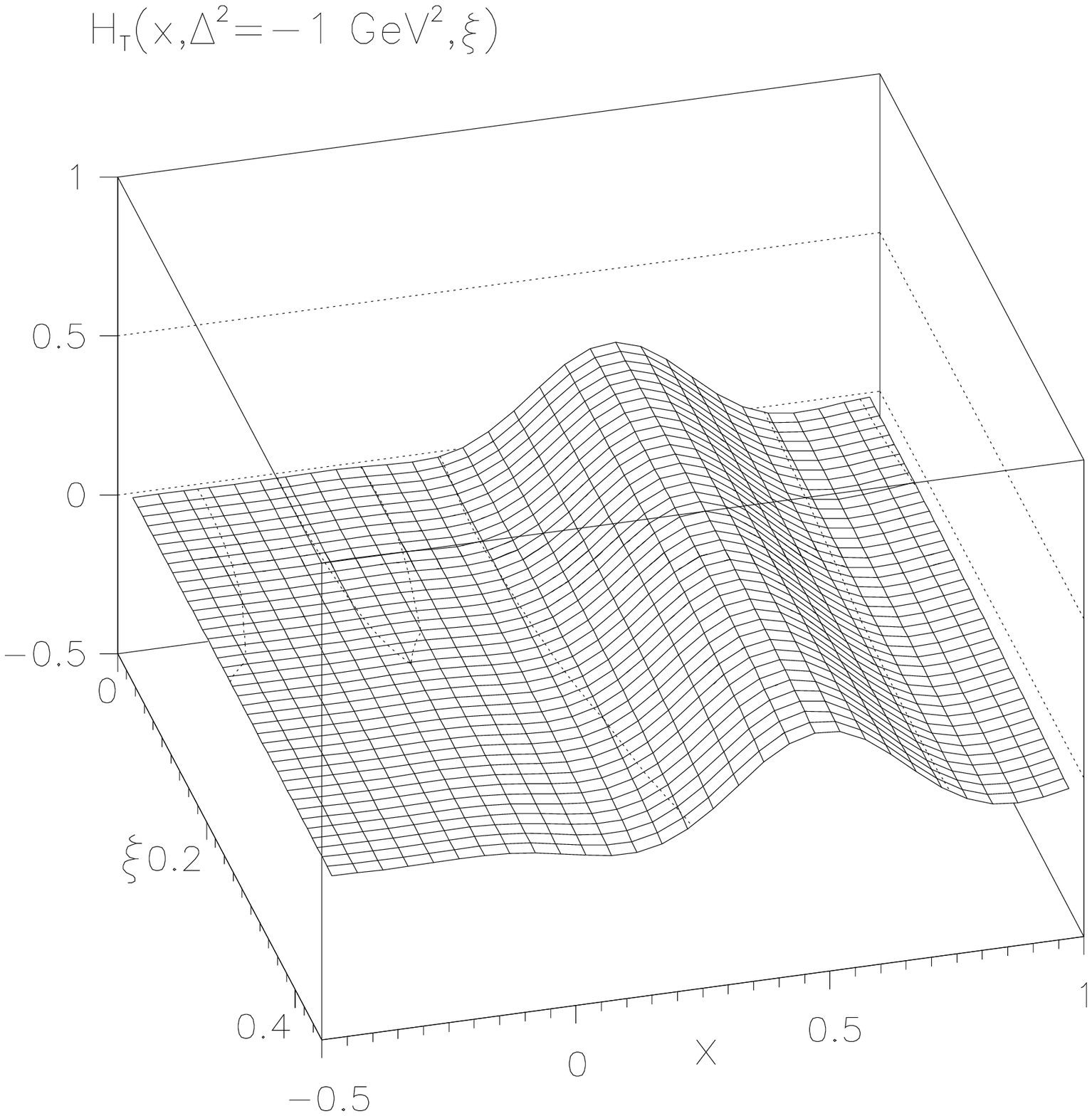}
\vspace{12.0cm}
\caption{}

\end{figure}

\newpage

\begin{figure}[h]
\vspace{12.cm}
\includegraphics{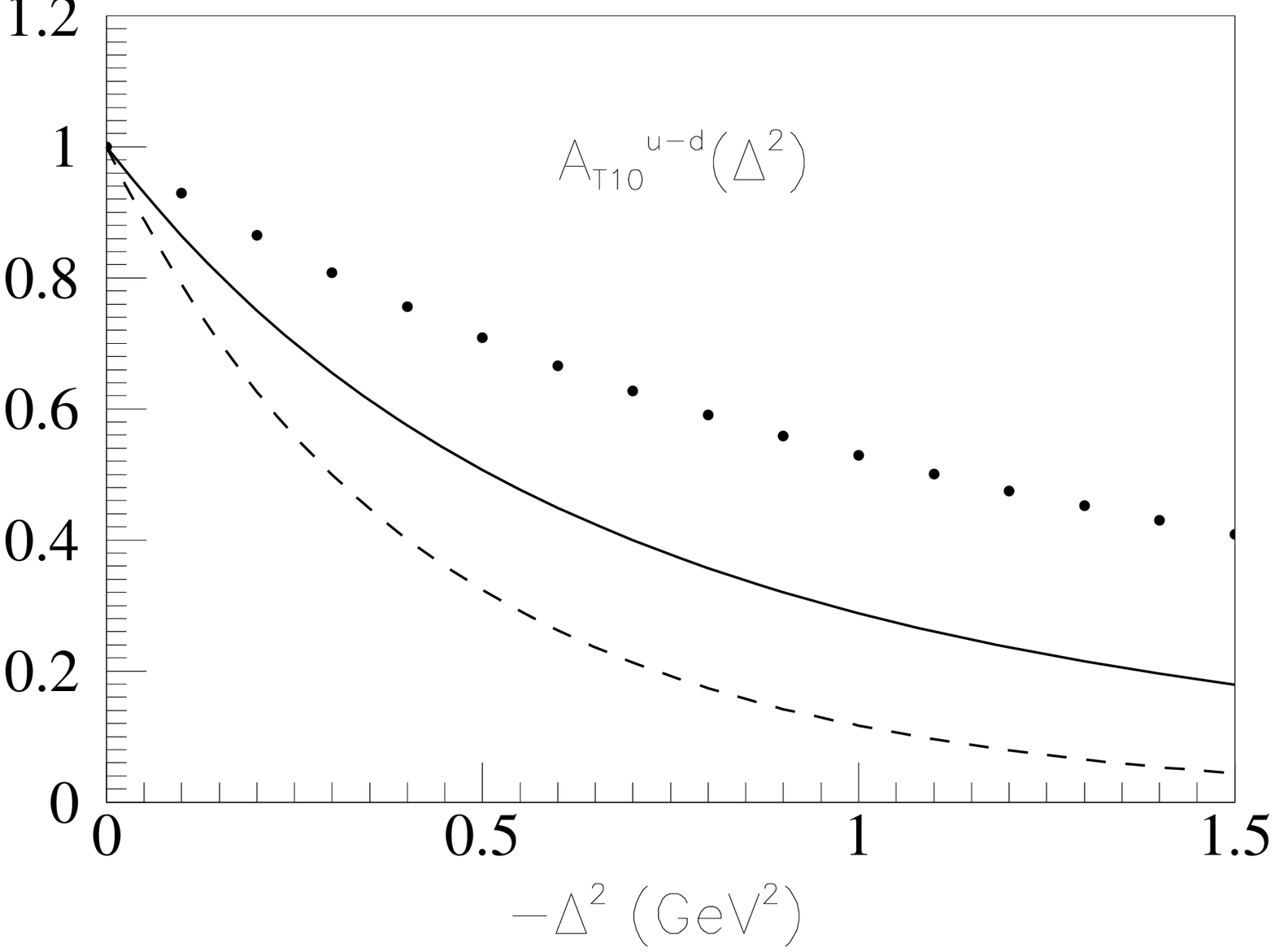}
\caption{}
\end{figure}

%\newpage
%
%\begin{figure}[h]
%\vspace{12.cm}
%\special{psfile=xg1p_ik.ps hoffset=80 voffset=-500 hscale= 55 vscale=55}
%\caption{}
%\end{figure}

%\newpage

%\begin{figure}[h]
%\vspace{8.2cm}
%\special{psfile=xg1n_ik.ps hoffset=80 voffset=-500 hscale= 55 vscale=55}
%\caption{}
%\end{figure}

%\newpage

%\begin{figure}[h]
%\vspace{8.2cm}
%\special{psfile=fig7.eps hoffset=80 voffset=-500 hscale= 55 vscale=55}
%\caption{}
%\end{figure}


\begin{thebibliography}{99}
\bibitem{trans} 
V. Barone, A. Drago, P.G. Ratcliffe, Phys. Rept. 359, 1, (2002);
V. Barone, P.G. Ratcliffe, ``Transverse spin physics'',
River Edge, USA, World Scientific (2003).
\bibitem{first} D. M\"uller, D. Robaschik, B. Geyer, F.M. Dittes,
and J. Ho\v{r}ej\v{s}i, Fortsch. Phys. 42, 101 (1994);
A. Radyushkin, Phys. Lett. B 385, 333 (1996);
X. Ji, Phys. Rev. Lett. 78, 610 (1997).
\bibitem{md}
K. Goeke, M.V. Polyakov, and M. Vanderhaeghen,
Prog. Part. Nucl. Phys. 47, 401 (2001); 
M. Diehl, Phys. Rept. 388, 41 (2003);
X. Ji, Ann. Rev. Nucl. Part. Sci.54, 413 (2004);
A.V. Belitsky and A.V. Radyushkin, Phys. Rept. 418 (2005).
\bibitem{hod} P. Hoodbhoy and X. Ji, Phys. Rev. D58, 054006 (1998).
\bibitem{die1} M. Diehl, Eur. Phys. J. C 19, 485 (2001).
\bibitem{bmk} A.V. Belitsky, D. M\"uller, and A. Kirchner,
Nucl. Phys. B629, 323 (2002).
\bibitem{pi1} M. Diehl, T. Gousset, and B. Pire,
Phys. Rev. D 59, 034023 (1999); J.C. Collins and M. Diehl, Phys. Rev. D 61,
114015 (2000).
\bibitem{pire} D.Yu. Ivanov, B. Pire, L. Szymanowski, and O.V. Teryaev, 
Phys. Lett. B550, 65-76 (2002); Phys. Part. Nucl. 35, 67 (2004).
\bibitem{lat} M. G\"ockeler at al. (QCDSF Collaboration),
Nucl. Phys. A755, 537 (2005); Phys. Lett. B627, 113 (2005).
\bibitem{goe} M. Kirch, P.V. Pobylitsa, and K. Goeke, 
Phys. Rev. D72, 054019 (2005).
\bibitem{die2} M. Diehl and Ph. H\"agler, hep-ph/0504175;
M. Burkardt, hep-ph/0505189.
\bibitem{MIT}
A. Chodos, R. L. Jaffe, K. Johnson, C. B. Thorn and V. Weisskopf,
Phys. Rev. D 9, 3471 (1974).
\bibitem{ji1} X. Ji, W. Melnitchouk, and X. Song,
Phys. Rev. D 56, 5511 (1997).
\bibitem{jaf} R.L. Jaffe, Phys. Rev. D 11, 1953 (1975).
\bibitem{jija1} R.L. Jaffe and X. Ji, Phys. Rev. D 43, 724 (1991).
\bibitem{jija2} R.L. Jaffe and X. Ji, Phys. Rev. Lett. 67, 552 (1991);
Nucl. Phys. B 375, 527 (1992).
\bibitem{strat} M. Stratmann, Z. Phys. C 60, 763 (1993);
S. Scopetta and V. Vento, Phys. Lett. B 424, 25 (1998).
\bibitem{sv2} S. Scopetta and V. Vento, Phys. Lett. B 460, 8 (1999), 
Erratum-ibid. B 474, 235 (2000).
\bibitem{dan}  I.V. Anikin, D. Binosi, R. Medrano, S. Noguera, and V. Vento,
Eur. Phys. J. A14, 95 (2002).
\bibitem{BETZ}
X. Song and J. S. McCarthy,
Phys. Rev. C 46, 1077 (1992);
M. Betz and R. Goldflam,
Phys. Rev. D 28, 2848 (1983).

\end{thebibliography}
\end{document}